# Solar Wind Electric Fields in the Ion Cyclotron Frequency Range

P. J. Kellogg (1),  S. D. Bale (2), F. S. Mozer (2), T. S. Horbury (3), H. Reme (4)

## 1. INTRODUCTION

In earlier papers (Kellogg and Lin 1997, Kellogg 2000, Kellogg et al 2001) which attempted to understand what effects were responsible for the violation of the conservation of magnetic moment in the solar wind it was pointed out that electric fields in the range which would resonate with the ion cyclotron frequency have not been adequately investigated.  In most spinning spacecraft, fields in this frequency range are overwhelmed by the photoelectric variation of the antenna potentials.

More generally, it is expected that there are fluctuating forces which  replace collisions, not only to maintain isotropy, but also to make valid the assumptions leading to MHD, since that theory seems to work in the solar wind.  The violation of conservation of magnetic moment is a convenient measure of these effects which can be treated quantitatively, but is not the only one.

A first attempt at measurement of electric fields in the relevant frequency range was made with the URAP instrument on Ulysses.  The spin axis of Ulysses was pointed near the sun, reducing photoelectric variation, but the instrument used a digital transform for on-board frequency analyses whose leakage from the spin frequency was so high that only occasionally were signals in the relevant frequency range visible above this unwanted signal (Lin et al 2003) .

The RPWS experiment on the Cassini spacecraft seemed to offer an excellent opportunity as Cassini is three-axis stabilized (Kellogg et al, 2001). However the measurements seemed to be contaminated by an instability on the spacecraft wake (Kellogg et al, 2003), and it was uncertain how much of the observed signal represented the undisturbed solar wind.

The present work analyzes data from the Cluster satellites.  These are spinning. However the antennas consist of small spheres.  These are on the ends of wires which are not part of the sensor, and which are biased by a current which nearly cancels the photoelectron emission.   These antennas have much reduced photoelectric variation which has allowed the observations of low frequency electric fields presented here.

## 2. THE  MEASUREMENTS

The Cluster satellites, a set of four nearly identical spacecraft which are programmed to fly close together, carry two orthogonal pairs of double probes for the measurement of electric fields. (Gustafsson et al 1997, 2001)   The Cluster mission is primarily aimed at the study of the magnetosphere and its boundary layers, but occasionally the satellites find themselves in the solar wind.  The position of the Cluster spacecraft during the period which has been analyzed here is shown in Fig. 1, together with a model of a typical Earth's bow shock.  A line in the direction of the magnetic field is shown for a certain time (0322 UT). Note however that the shock did not stay in this model position during the whole day. Some solar wind conditions will be shown later, in Fig. 7. The spacecraft were in the free solar wind, unconnected by magnetic field to the bow shock for the periods which were used here, but for some short periods there were magnetic connections to the bow shock, and for the period 0345-0600, the solar wind dynamic pressure was so low that the bow shock expanded sunward of the spacecraft. All of these periods have been excluded.

In Fig. 2 are shown four examples of the electric fields as seen from the probe 1-probe 2 potential difference (in red), and $\mathbf{V_{sw}} \times \mathbf{B}$ projected onto the same probe direction. The electric field, thus, is nearly the raw data, and has been minimally manipulated. In these plots, the electric field is sampled at 25 samples per second, and the magnetic field is sampled at a comparable rate, approximately 22 samples per second (from the FGM experiment, Balogh et al 1997, 2001). The solar wind speed, from the CIS instrument, is sampled only at four second intervals (Reme et al 1997, 2001).

Several things can be seen from these figures, which show that, overall, the raw electric field measurements track the $\mathbf{V_{sw}} \times \mathbf{B}$ measurements very well, including the abrupt change shown in (d). First, there are fluctuations in the electric field which are not present at the same amplitude and time scale in $\mathbf{V_{sw}} \times \mathbf{B}$. In this, it might be argued that $\mathbf{V_{sw}}$ is not sampled as rapidly as E, but B is sampled at approximately the same cadence as E, and the fluctuations of $V_{sw}$, given that the fluctuations of B are partly Alfvenic, ought to contribute only in the ratio $V_A / V_{sw}$ to the electric fluctuations.

Second, large spikes are seen in some of the electric field plots. These are thought to be due to the probes encountering the wake of the main body of the spacecraft. These "spin tones" contribute considerably to the electric field spectrum after it has been transformed to the GSE coordinate system. They vary considerably in amplitude, being sometimes nearly absent. Since they vary only slowly from one rotation to the next, they mainly contribute harmonics of the spin frequency. In order to minimize leakage from these very strong harmonics, the Fourier analyses presented below are done on 4096 sample sets. 4096 samples is 41.005 periods of the spin, and this being nearly an integral number of spins, leakage is minimized.

Third, it will be seen that sometimes the amplitude of E is greater, and sometimes less, than the amplitude of $\mathbf{V_{sw}} \times \mathbf{B}$. Partly this is due to errors in the three measurements, but some of it is due to a dependence of the measurement of electric field on the ambient electron density. The bias current of the probes is adjusted so that the probes are slightly positive with respect to the plasma. This is done because the probe-plasma impedance depends strongly on whether ions or electrons are being collected, and it is safer to maintain the probes in the electron collection region. As a consequence, however, the potential difference between two probes is slightly less than the full electric field would give. This difference depends on the plasma density. There is also an effect of the conducting cables which lead to the probes which will reduce the potential differences. In Fig. 3 is shown a comparison between the amplitude of the projected $\mathbf{V_{sw}} \times \mathbf{B}$ and of the electric field as a function of average probe voltage, which is a measure of electron density. (Pedersen, 1995). The line shown is our fit to these effects, and it has been used to correct the electron density effect on the measured fields in what follows.

## 3. ELECTRIC FIELDS

In order to determine the effect of these electric fields on the ions of the solar wind it is necessary to determine the electric field in the average frame of the ions. The solar wind also has large magnetic fluctuations. These are Lorentz transformed by the solar wind velocity to give an apparent electric field, $\mathbf{V_{sw}} \times \mathbf{B}(f)$. Therefore it is necessary to subtract $\mathbf{V_{sw}} \times \mathbf{B}(f)$ from the measured electric field. This is least important for the X (GSE) component of the fields, since the X component of $\mathbf{V_{sw}} \times \mathbf{B}$ does not involve the X component, the largest component, of the solar wind velocity. In Fig. 4 are shown the power spectra of $\mathbf{E_x}$, of $\mathbf{E_x} + (\mathbf{V_{sw}} \times \mathbf{B})_x$ and of $(\mathbf{V_{sw}} \times \mathbf{B})_x$. In order to calculate these spectra, the solar wind velocity and the magnetic fields have been interpolated to the sample times of the electric field. As said above, this is not a serious concern for the magnetic field which is sampled at approximately the same cadence as the electric field, but does not supply velocity values in the higher frequency range. It is argued above that this does not seriously affect the results.

As expected, $(\mathbf{V_{sw}} \times \mathbf{B})_x$ does not make up a large part of the measured component of $\mathbf{E_x}$. These measurements, therefore, do not require a large correction due to $\mathbf{V_{sw}} \times \mathbf{B}$. The case is somewhat different for $\mathbf{E_y}$. Fig. 5 shows the same sorts of spectra for the GSE Y components. Here $\mathbf{E_y}$ and $(\mathbf{V_{sw}} \times \mathbf{B})_y$ are of the same order, and the remaining $(\mathbf{E} + \mathbf{V_{sw}} \times \mathbf{B})_y$ is perhaps affected by the errors in the measurements. However, the corrected spectrum is slightly smaller than that of $\mathbf{E_x}$ and is probably at least roughly correct. The X component is therefore the more accurate measure of the electric field. The power spectrum of $(\mathbf{E} + \mathbf{V_{sw}} \times \mathbf{B})_x$ is fitted by the sum of

$$10^{-4} f^{-5/3} \; (mV/m)^2/Hz \quad \text{(power law part)}$$

and

$$2. \; 10^{-4} \; (mV/m)^2/Hz. \quad \text{(plateau)}$$

Bale et al (2005) have given a more thorough discussion of the spectrum, and have shown that there is a region above the power law which fits the dispersion relation for kinetic Alfven waves, an attractive possibility especially from the point of view of identification of these fluctuations. The reader is directed to that paper for such a discussion. Here we are more concerned with the effects of these fields than with their identification.

In Fig. 6, these measurements of the power spectrum of the X component of the electric field are compared with Ulysses measurements at about the same distance from the sun (1.34 AU) As the noise threshold for the Ulysses experiment was comparable to average electric fields, the spectra shown, averages over 10-20 min, are chosen from the large signals. The measured electric field spectra do not match well with Ulysses measurements made in the frequency range 9-484 Hz (Stone et al 1992, Kellogg 2000, Lin et al 2003). Therefore there is substantial concern as to the electric fields above about 1 Hz, where the Cluster spectra become nearly flat. In particular, there was concern that the spectra are affected by leakage from the strong spin harmonic tones. As a check the spectra were calculated both with and without a Hamming window. The windowing reduces the heights of the spin tone peaks and increases their widths slightly, of course, but has no discernable effect on values well separated from the harmonic frequencies. Furthermore, and most importantly, the fluctuations at frequencies well above 1 Hz can already be seen in the time series of Fig. 2, which cannot be affected by leakage. Hence this concern is unfounded.

Another source of error might have been a difference in the response of the probes to density fluctuations. A difference in the probe response would contribute a difference signal, which would be interpreted as an electric field. Density fluctuations are large, and have a roughly $f^{-4/3}$ spectrum, or at least a spectrum which is not flat (Celnikier et al, 1983, 1987, Kellogg et al, submitted to Annales Geophys, 2005). The spurious electric field produced would also have the same power law, and so could not produce the flat spectrum which is observed. Furthermore, calibrations show that there is essentially no difference between probes 1 and 2, and no significant difference in the slopes of the responses of all probes. So this cannot account for the difference. On the other hand, the Ulysses measurements were obtained with long antennas (two 35 m monopoles) made of a thin copper-beryllium tape. There is always a question as to the effective length of such antennas. The measurements shown are assuming an effective length of 23 m, a length which is appropriate for the capacitively coupled high frequency region, but accurate calibrations for the resistively coupled, low frequency region have not been made. It seems highly unlikely, however, that the effective length could be short enough to explain the difference. The question of the disagreement with Ulysses measurements remains unresolved, but we cannot find a reason for discarding the Cluster measurements. However, instabilities of a plasma flowing around an antenna have been observed in the past, (Neubert et al 1986, Kellogg et al 1990, Kellogg et al 2003). It might possibly be that the "plateau" is such an instability.

The $E_x$ field in the plateau region, if it is real, is sufficiently larger than $(\mathbf{V} \times \mathbf{B})_x$ that no correction for it is necessary.

The spectra shown in Figs. 4 and 5 are averages over the period 0003-0727 UT on 19 Feb, 2005, with the exclusions noted above. In Fig. 7 are shown some representative time behaviors of the electric fields, and basic plasma parameters for comparison. The lowest panels give the solar wind solar wind speed (km/s) and the plasma density. Regions of reduced speed are regions where the bow shock has expanded over the spacecraft. In the region 0400-0420 UT, the spacecraft were magnetically connected to the bow shock and that is why those data are excluded. The low frequency behavior, .007 Hz to .2 Hz, is usually a power law at $f^{-5/3}$ and the coefficient of this law has been plotted in the topmost panel. We have not been successful in removing spin tones from a calculation of the power in the plateau region, so the next two panels show the total power in two regions between peaks. It will be seen the the power law amplitude is quite variable, as is expected for most solar wind fluctuations, and that there seems to be a tendency for large amplitudes in the places where the plasma density is changing or has a discontinuity. The power in the .77-.95 Hz region shows rather little variation, which adds to our suspicion of these data. On the other hand, the 1.5-2 Hz region shows fluctuations which can also be seen on individual spectra (not shown).

4. DIFFUSION IN VELOCITY SPACE.

The solar wind seems to behave as if it were a collisional plasma, though collision rates are quite small. It is generally thought that wave-particle interactions replace collisions in such cases. Such effective collisions are necessary for the validity of MHD theory. In this section, however, we consider the observed near isotropy of the observed proton velocity distributions (Marsch et al 1983, Marsch 1991), which lends itself to a quantitative discussion. By near isotropy, we mean that the observed ratios of $T_{par}$ to $T_{perp}$ are within a factor of 2 of unity, whereas values of 100 to 1000 would be expected with conservation of magnetic moment where particle-particle collisions are as rare as they are in the solar wind. (Lemaire and Scherer 1973)

Major progress toward understanding the cause of near isotropy has made by Gary et al (2001), and by Kasper et al (2002). Gary et al showed that the electromagnetic proton cyclotron instability seems to limit large Tperp/Tpar while Kasper et al showed that the resonant proton firehose instability limits larger values of Tpar/Tperp when β is larger than about 1. There remains a puzzle when Tpar/Tperp is larger than 1, as is expected for conservation of magnetic moment, but β is small. This is the interesting case closer to the sun where β is typically smaller, and is especially relevant to the Helios results (Marsch et al 1983).

Although it is believed that wave-particle interactions must provide effective collisions, it has not been certain just what waves were involved. In earlier work, one of the authors (Kellogg and Lin 1997, Kellogg 2000 ) assumed that the principal contribution to effective collisions would be waves in the frequency range resonant with ions at their cyclotron frequency, which would be Doppler shifted to the vicinity of 1 Hz. Here we see that there is a large part of the electric field spectrum at lower frequencies, and it is necessary to consider the effect of this part also.

Kennel and Engelmann (1966) and Ichimaru (1973) have given diffusion rates for the perpendicular component of velocity in terms of electrostatic fields. The Ichimaru formulation seems to be based on having Fourier amplitudes which are stable for a long time, while the Kennel and Engelmann formulation assumes unstable waves. The Ichimaru formulation implies that only those waves which are exactly resonant with a harmonic of the cyclotron frequency of the particle have any effect. In the case considered here, it would seem that any of this turbulent mix of electric fields ought to have some effect to change the velocities of the particles. Further, the diffusion rates of both formulations depend on the four dimensional distribution function of $E^2$ in ω,**k** space, and here only one dimension, considered to be one component of **k**, is measured. It would be necessary, therefore, to make some only weakly justified assumptions about the behavior of the distribution function in the other three dimensions. Hence a different approach is used here, more akin to the rough calculation of Kellogg and Lin (1997 ) and of Kellogg (2000). The fluctuations are represented by short sinusoidal packets. The idea is that the fluctuations are subject to nonlinear effects like mode conversion which change them rapidly, and that we can represent such effects phenomenologically by short packets without having to treat the nonlinear effects in detail.

Let a particle be subject to a short packet of exactly N cycles of electric field in the x direction and to a static magnetic field B in the z direction. N is to be a small integer, and is, clearly, a measure of coherence.

$$E(t) = E_0 \sin \omega t \qquad \text{for } \omega t = 0 \text{ to } 2N\pi \qquad (1)$$

The solution of the equations of motion, in the rest frame of the plasma, and for a particle with velocity $v_{x0}, v_{y0}$ at $t = 0$ is:

$$v_x = -V_E \sin \omega t + [v_{x0} + V_E] \cos \Omega t + v_{y0} \sin \Omega t$$

$$v_y = v_{y0} \cos \Omega t - [v_{x0} + V_E] \sin \Omega t \qquad (2)$$

$$\text{with } V_E = (q E_0/m)/[\omega/(\omega^2 - \Omega^2)] \qquad \text{and } \Omega = qB/m \qquad (3)$$

Although the terms in $v_{xo}$ and $v_{yo}$ are larger than the terms in $V_E$ in the solar wind, they average to zero over many packets, so they are neglected. The change in perpendicular velocity is then:

$$\Delta V_\perp^2 = \left(v_x^2(t_f) + v_y^2(t_f) - v_{x0}^2 - v_{y0}^2\right) = 2V_E^2\left(1 - \cos\left(2\pi N \frac{\Omega}{\omega}\right)\right) \qquad (4)$$

Here $t_f = 2N\pi/\omega$ is the temporal length of the packet.
Averaging for a time T over many packets at different frequencies gives:

$$\frac{\delta V_\perp^2}{T} = \left(\frac{q}{m}\right)^2 \int 2E_0^2(f) \left[\frac{\omega}{\omega^2 - \Omega^2}\right]^2 \left[1 - \cos\left(2\pi N \frac{\Omega}{\omega}\right)\right] df \qquad (5)$$

with $\omega = 2\pi f$ and $E_0$ of eq.(1) is now a function of frequency $E_0(f)$.

The integrand remains finite at $\omega = \Omega$ as will be shown in Fig. 8.

The observed electric field spectrum is a power law of exponent $-5/3$ at low frequencies, so that the $\omega^2$ dependence in $V_E^2$ implies that frequencies in the resonant range are still the most important in spite of the rising of the spectrum toward low frequencies.

The connection to the power spectrum is not quite immediate. As the duration of a packet is $2N\pi/\omega = N/f$, each contributes $1/2\, E_0^2\, N/f$ to the time integral of $E^2$. If, in time T, there are $n(f)\, T\, df$ of these packets with frequency between $f$ and $f + df$, ($\omega = 2\pi f$), then their contribution into the power spectrum gives:

$$\int_o^T E^2(t)\, dt = \int E^2(f)\, df = \int \frac{E_o^2}{2}\frac{N}{f} n(f)\, df \qquad (6)$$

where the first step is Parseval's theorem. Then $E^2(f) = 1/2\, E_0^2\, (N/f)\, n(f)$, and the diffusion in time T becomes:

$$\frac{\delta V_\perp^2}{T} = \left(\frac{q}{m}\right)^2 \int \frac{4f}{N} E^2(f) \left[\frac{\omega}{\omega^2 - \Omega^2}\right]^2 \left[1 - \cos\left(2\pi N \frac{\Omega}{\omega}\right)\right] df \qquad (7)$$

again with $\omega = 2\pi f$. The fundamental parameters q, m, and $E^2(f)$ are the same as in Ichimaru (1973) and Kennel and Engelmann (1966) but the expression differs in neglecting some Bessel functions which provide a correction for finite wavelength of the packet, and in being averaged over the particle velocity instead of being valid only for a specific resonant velocity.

In the above equation, $E^2(f)$ is measured in the proton rest frame. The observed frequency is presumed to be Doppler shifted by the solar wind speed. It is expected that the fluctuations obey a dispersion relation like $\omega = k\, V_W$ as this is true for all MHD waves at low frequency, and that $V_W$ is much less than $V_{sw}$. The relation between the frequency, f, of the waves in the proton rest frame and the observed frequency is then $f = (V_W/V_{sw})\, f_{obs}$, with the wave speed, $V_W$, being either the ion sound speed or the Alfven speed. 50 km/s is a typical value for wave speed, giving a factor of 1/7 here. Hence $E^2(f)$ in the equations above should be replaced by $E_{obs}^2((V_{sw}/V_w)\, f)$. In Fig. 8 upper panel the coefficient of $E_{obs}^2$ in the integrand is plotted for N = 1, 2 and 3. The peaks are in that order, that is, N = 3 gives the highest and most narrow peak. The cyclotron frequency in the proton rest frame is indicated by a dashed line. It will be seen that the original assumption of Kellogg and Lin, 1997, that the main contribution comes from the region around the proton cyclotron frequency, is born out in this analysis. In the lower panel, the whole integrand including $E^2(f)$ from eq. (7) and corrected for Doppler shift, is plotted for N = 2, and for the two forms of the spectrum. Obviously, the higher values are due to the inclusion of the plateau part of the spectrum.

During the periods under analysis, the solar wind speed was fairly constant at 350 km/s, the magnetic field fairly constant at 10 nT and assumed to vary with distance from the sun as $r^{-1.5}$, a compromise between $r^{-2}$ for the radial component of B and $r^{-1}$ for the transverse component. The proton temperature was generally around .2 MK, Using these values and $E^2(f)$ from Fig. 4, the diffusion calculated in eq. (7) is evaluated at:

$$\frac{dV_\perp^2}{dt} = 2.8\ 10^6 \frac{m^2}{s^3} \quad Power\ Law \qquad (8a)$$

$$8.9\ 10^6 \frac{m^2}{s^3} \quad with\ plateau \qquad (8b)$$

The development leading to Eq. (7) was begun with the thought that the spectrum rising toward low frequencies might give a dominant contribution. This turned out not to be the case. Instead, the contribution near resonance is still dominant, and also dominates the contribution from higher frequencies.

In order to significantly violate conservation of magnetic moment, a diffusion as large as the change of perpendicular energy due to magnetic moment conservation is necessary. Conservation of magnetic moment would give a change of perpendicular velocity:

$$\frac{dV_\perp^2}{dt} = \frac{d}{dt}\left(\frac{V_{\perp o}^2}{B_O} B\right) = V_{\perp o}^2 \frac{V_{sw}}{B_O} \frac{dB}{dr} \tag{9}$$

$$= -1.1\ 10^4\ m^2/s^3$$

Clearly, the diffusion by electric fields is considerably larger than the minimum necessary to overcome conservations of magnetic moment.

## 5. CONCLUSIONS

Measurements of the electric field in the solar wind have been made, and corrected for the Lorentz transformation of magnetic fluctuations to show that there is a significant electric field in the range of the proton cyclotron frequency. The diffusion in velocity space due to these electric fields has been estimated, and found to be either $2.8\ 10^6$ or $9.\ 10^6\ m^2/s^3$ depending on which of two interpretations of the spectrum is correct. The decrease in the perpendicular energy, $W_\perp$ due to conservation of magnetic moment would be only about $1.1\ 10^4\ m^2/s^3$.

Consequently the diffusion due to electric fields is more than sufficient to maintain near isotropy of the proton velocity distributions.

It might also be that diffusion is caused by the magnetic fluctuations. This would be of the order of:

$$\left\langle (v_{th}\delta B)^2 \right\rangle \sim \frac{1}{50}\left\langle E^2 \right\rangle \tag{10}$$

and therefore much smaller than the diffusion due to electric fields. Here $E^2$ is taken from Figs. 4 and 6, and $\delta B^2$ from Fig. 5, using the value of the solar wind speed above. $v_{th}$ is the thermal velocity of the protons.

In the absence of electric field measurements, a great deal of work has been done on velocity space diffusion, or on perpendicular heating, by ion cyclotron waves (see Marsch et al 1982, Marsch and Tu 1997, Li et al 1999, Hollweg and Isenberg, 2002 and references therein). Assuming that a part of the magnetic spectrum is due to such waves, it is seen that diffusion due to magnetic fluctuations would also be sufficient to maintain isotropy. However, diffusion due to the electric fields is the dominant process, and therefore the isotropy of the proton distributions is mainly due to these electric fields.

It should be noted that this conclusion does not depend on the rough calculation of the diffusion coefficient leading to Eq. (7), but is only dependent on the measured spectra of E and B

ACKNOWLEDGEMENTS

Work at the University of Minnesota was supported by the U.S. National Aeronautics and Space Administration, contract NAS5-03076. Work at the University of California was supported by the U.S. National Aeronautics and Space Administration, grant NAS5-11944. Cluster work at Imperial College is supported by PPARC(UK). Work at CESR was supported by a grant from the Centre Nationale de Recherche Scientifique, France.

FIGURE CAPTIONS;

Fig. 1   Two views of the position of the Cluster spacecraft during these measurements, together with a standard model shock. The heavy line represents the orbit during these measurements and the light line is a typical direction of the magnetic field.

Fig. 2   Examples of the comparison of probe potential difference (red) and $\mathbf{V_{sw} \times B}$ projected onto the (spinning) interprobe direction.

Fig. 3   Ratio between the amplitudes of E and $V_{sw} \times B$ as a function of average probe voltage, a measure of plasma density.

Fig. 4   Power spectra of X (GSE) components of E, of $V_{sw} \times B$, and of $E + V_{sw} \times B$

Fig. 5   Power spectra of Y (GSE) components of E, of $V_{sw} \times B$, and of $E + V_{sw} \times B$

Fig. 6   Comparison of $E_x$ with Ulysses URAP measurements at 1.34 AU

Fig. 7   Time dependence of the power law part, and of two frequency ranges representative of the plateau, together with plasma parameters.

Fig. 8   Functions for calculation of the estimate of the diffusion coefficient. The dashed line indicates the proton cyclotron frequency.

REFERENCES:

Bale, S.D., et al, 2005, *Phys. Rev. Lett.* 94, 215002.

Balogh, A., et al 2001, *Annales Geophysicae* 19, no. 10-12:1207.

Balogh, A., et al 1997, *Space Science Reviews* 79, no. 1-2, 65.

Celnikier,L.M. et al 1983, *Astron.Astrophys.* **126**, 293.

Celnikier,L.M. et al 1987, *Astron. Astrophys..* 181, 138.

Gary,S.P. et al 2001 *Geophys.Res.Lett* **28,** 2759.

Gustafsson, G. et al 2001,*Annales Geophysicae* 19, no.12:1219-40.

Gustafsson, G. et al 1997, *Space Science Reviews* 79, no. 1-2:137-56.

Hollweg, J.W. & Isenberg, P.A. 2002 *J. Geophys. Res.* 107(A7),1147.

Ichimaru, S., "*Basic Principles of Plasma Physics*", W.A.Benjamin Inc., Reading MA., 1973, p 239

Kasper,J.C, et al 2002, *Geophys.Res.Lett* .**29**(17), 1839

Kellogg, P.J. et al 1990, *J.Geophys.Res*. 95, 7773.

Kellogg, P.J. & Lin, N., in SP-415, "*The 31st ESLAB Symposium, Correlated Phenomena at the Sun, the Heliosphere and in Geospace*" A. Wilson, Ed., ESTEC, Noordwijk, the Netherlands, 1997.

Kellogg, P.J. 2000, *Ap. J.* 528, 480.

Kellogg, P.J. et al, 2001, *Geophy.Res.Lett* 28, 87.

Kellogg, P. J. et al, 2003, *J. Geophys. Res., 108*(A1), 11.

Kennel,C.F. & Engelmann C.F. 1966, *Phys.Fluids* 9, 2377.

Lemaire,J. & Scherer,M. 1972, Rev. Geophys Sp. Phys. **11**, 427

Li.X., et al 1999, *J. Geophys. Res.* 104, 2521..

Lin, N. et al 2003, *Geophys.Res.Lett., 30*(19), 3-1.

Marsch,E, et al 1982, *J. Geophys.Res*. 87, 5030.

Marsch, E, & Tu, C-Y 1997, *Astron. Astrophys*. 319, L17

Neubert,T. et al, 1986, *Can. J. Phys*. 64, 1437.

Pedersen, A. 1995, *Annales Geophysicae*, 13(2), 118.

Reme, H. et al 1997, *Space Science Reviews,* 79(1-2), 303.

Reme, H. et al 2001 *Annales Geophysicae*, 19(10-12), 1303.

R.G. Stone et al 1992, *Ast. Astrophs. Suppl Ser*. 92, 291.

1. School of Physics and Astronomy, University of Minnesota, Minnneapolis, MN, USA,

2. Space Sciences Laboratory, University of California, Berkeley, CA. USA

3. Imperial College, London, England, UK

4. CESR, Toulouse, France

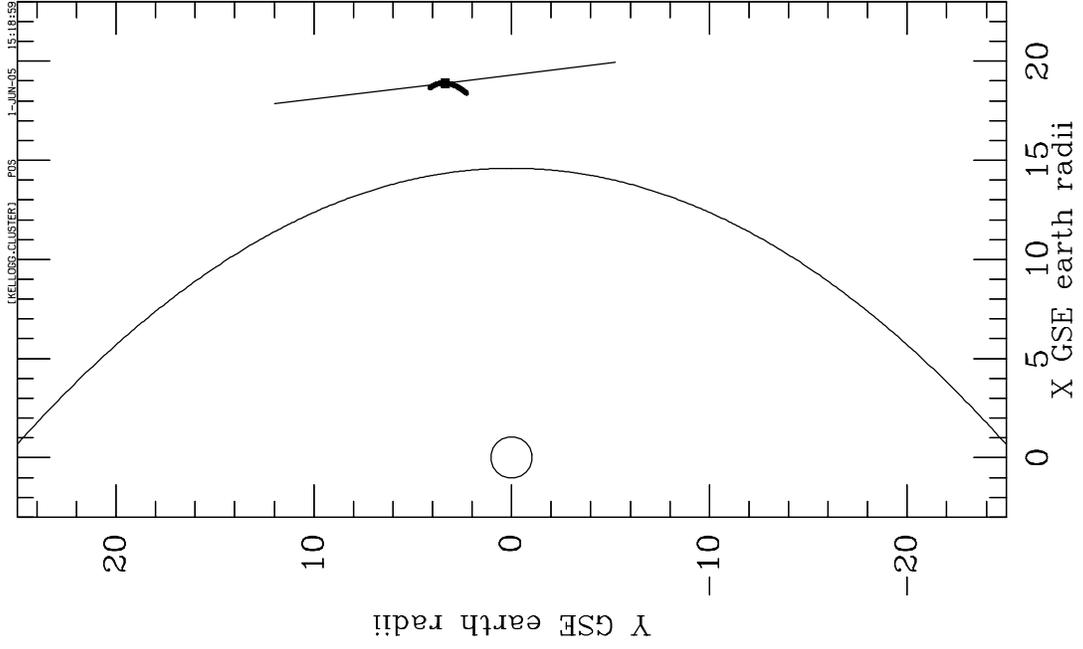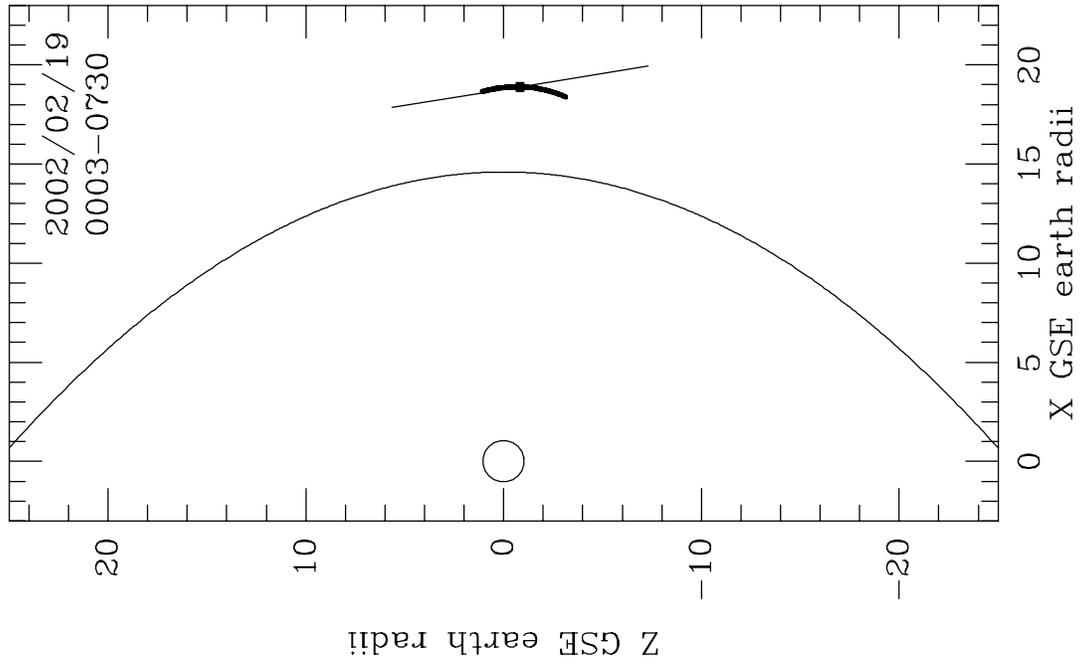

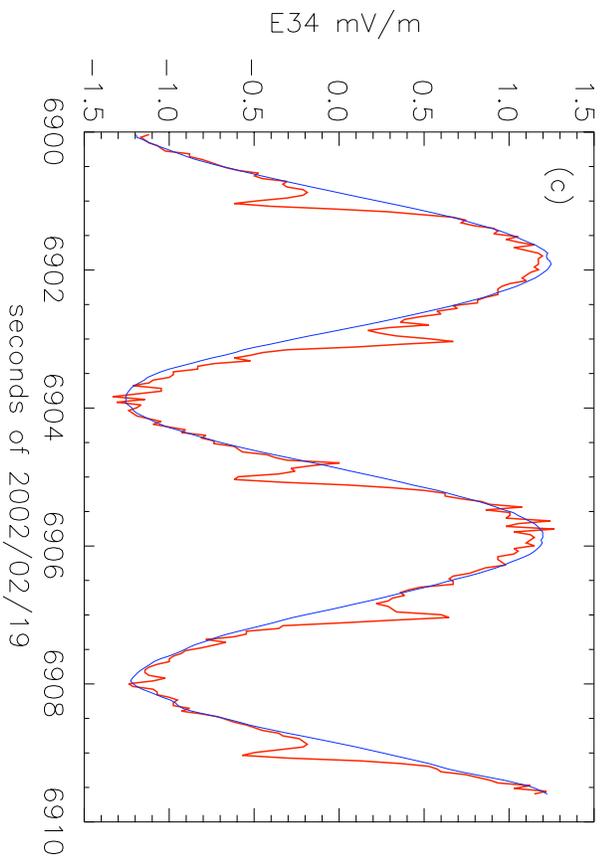
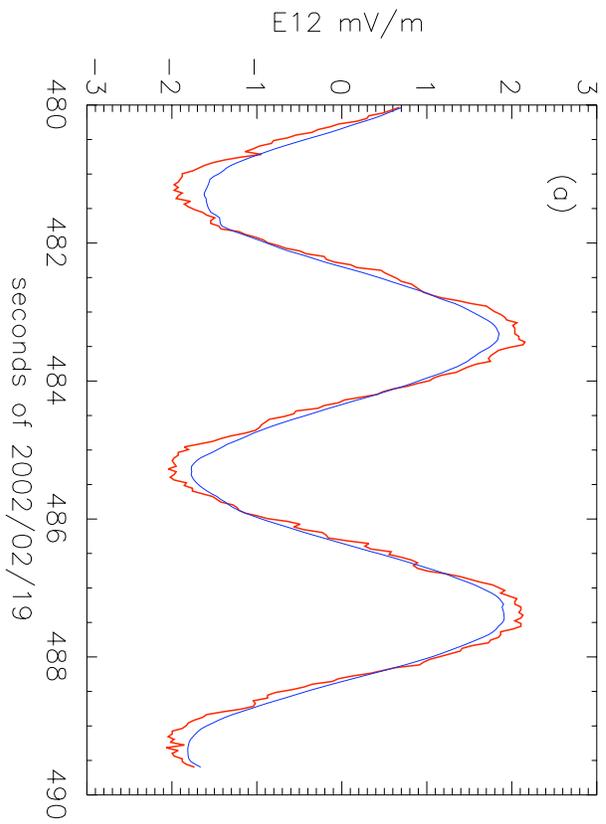
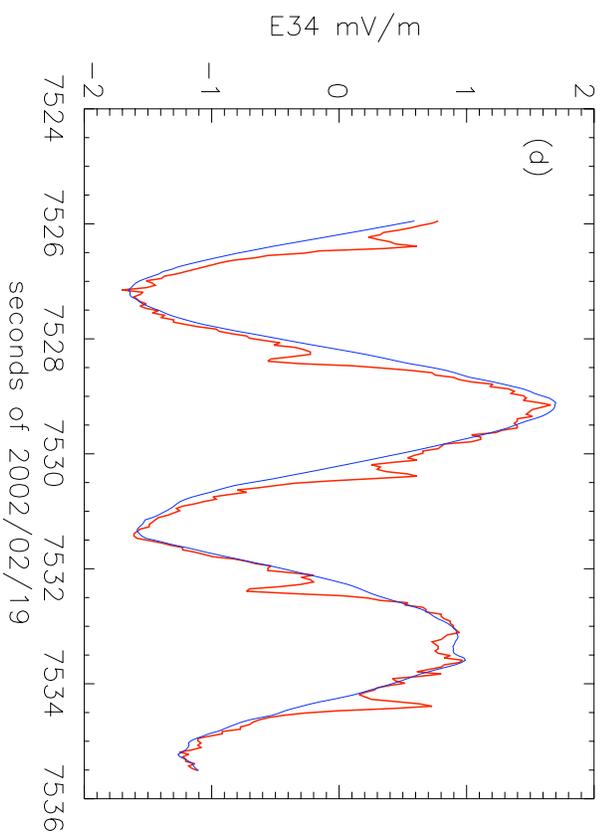
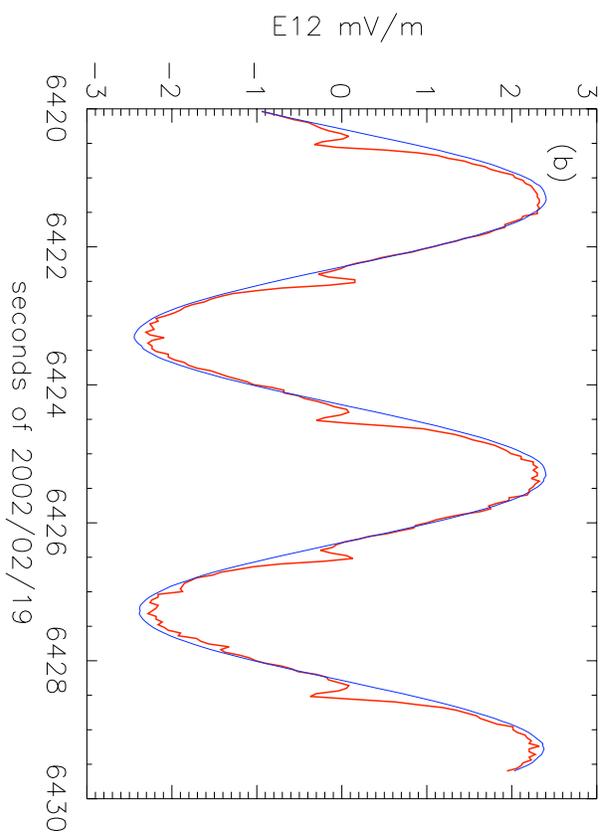

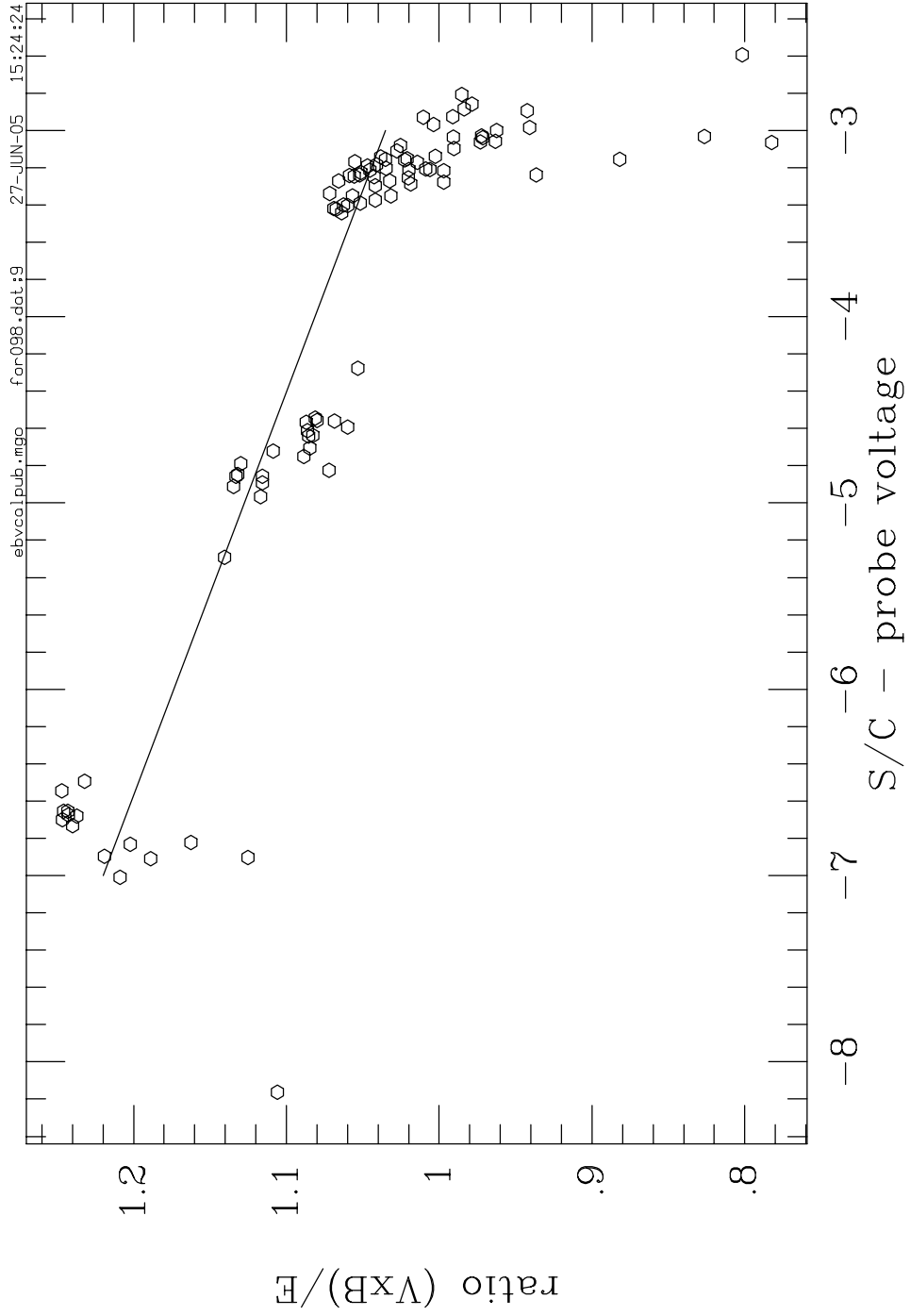

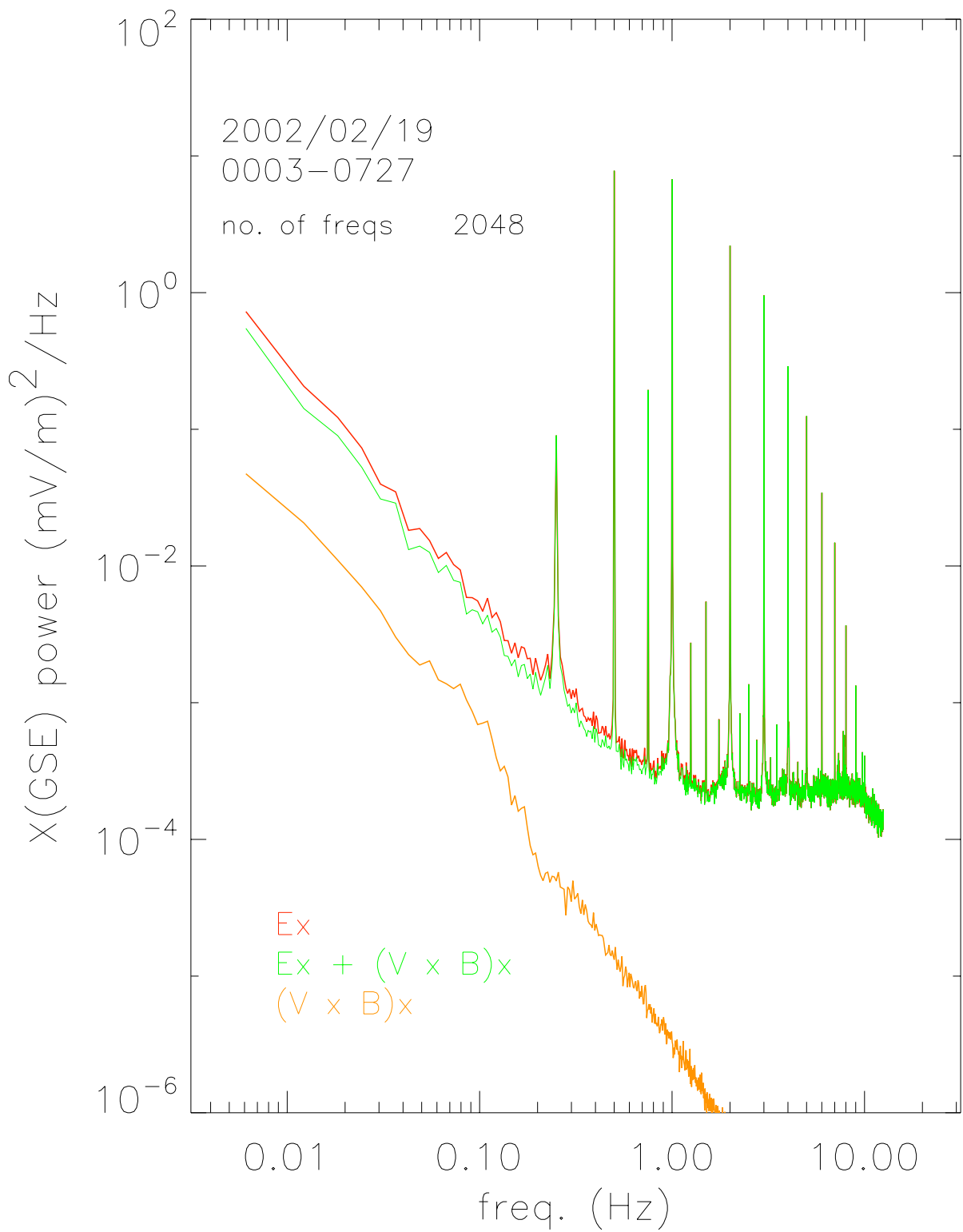

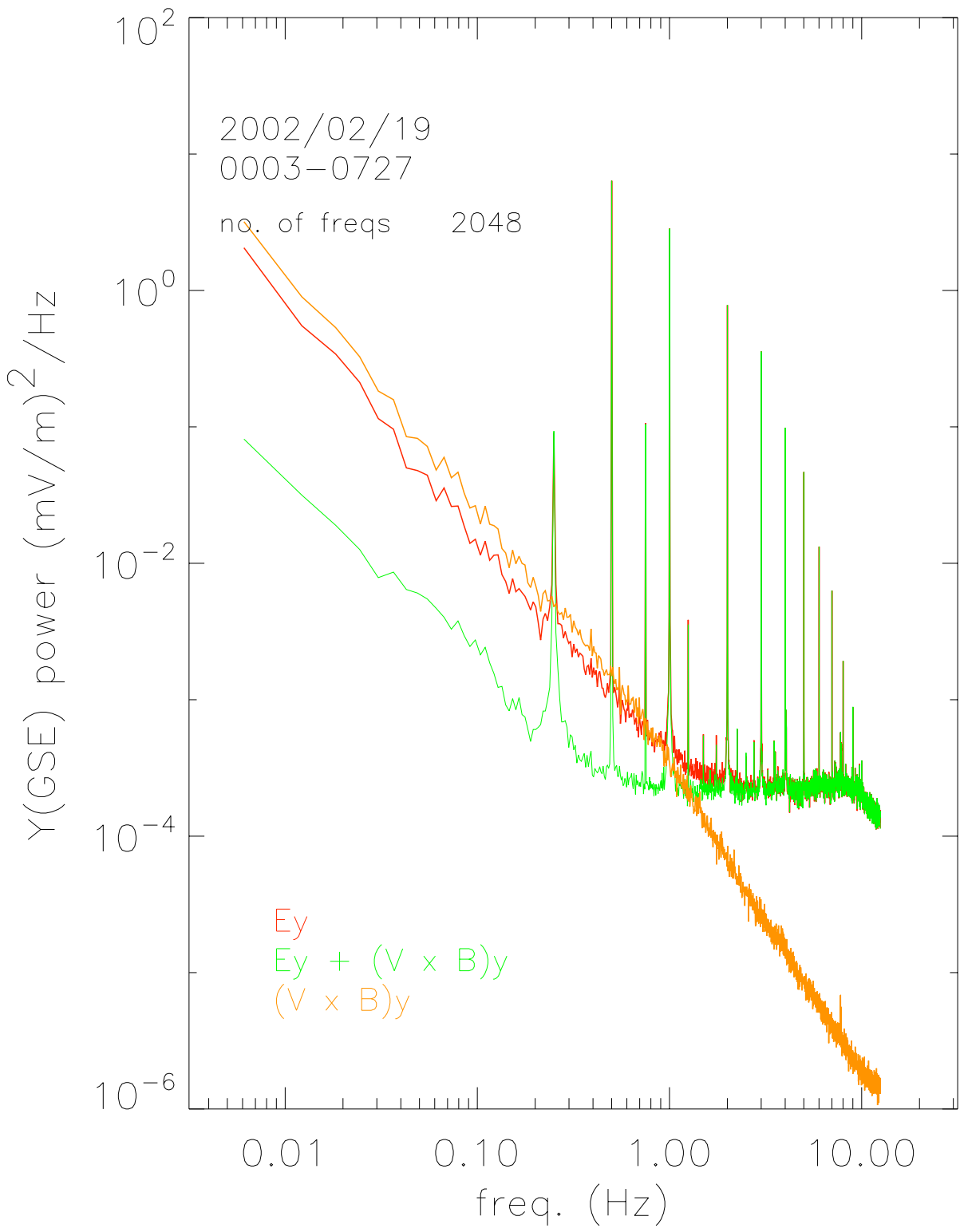

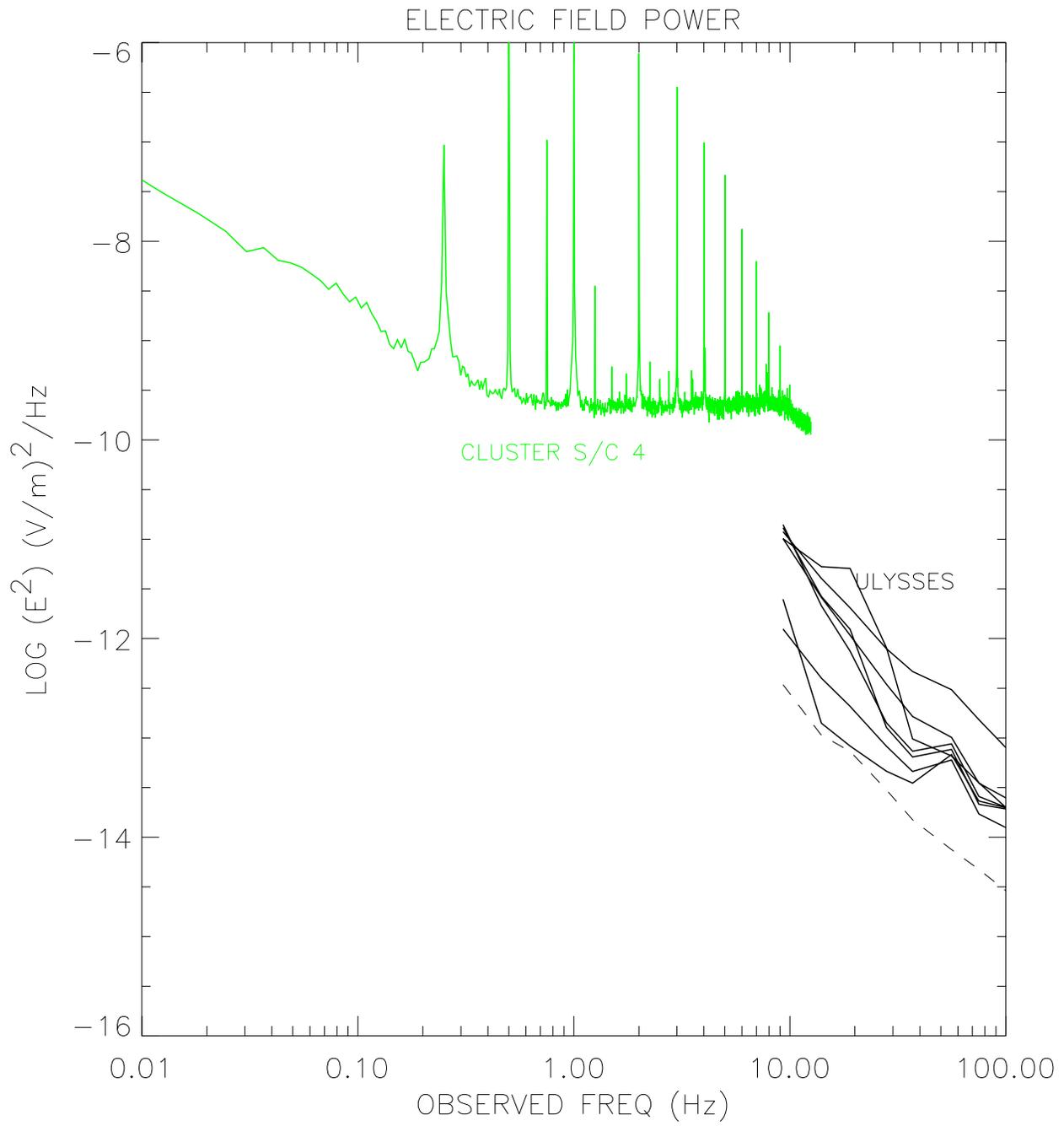

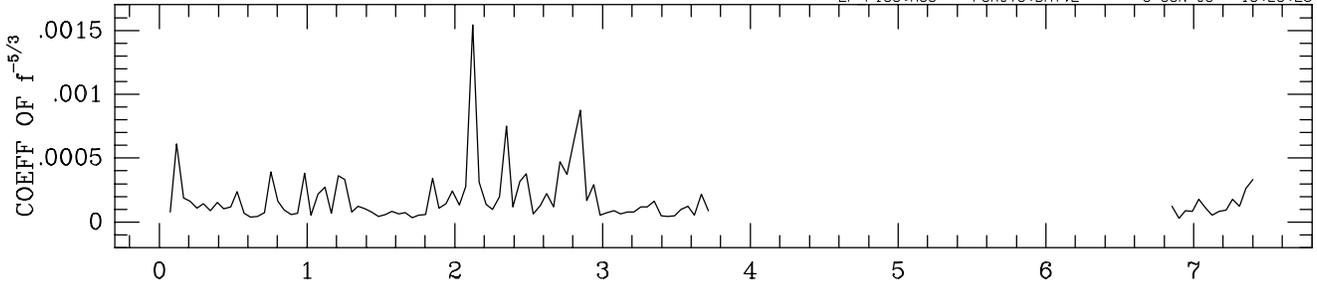
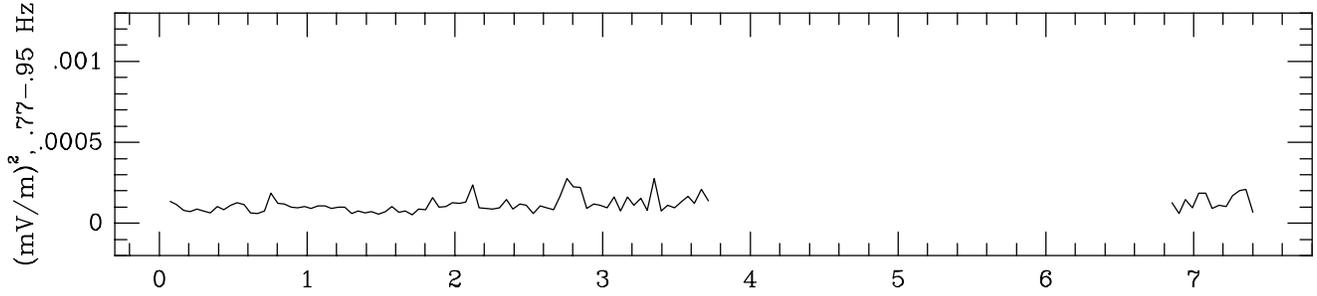
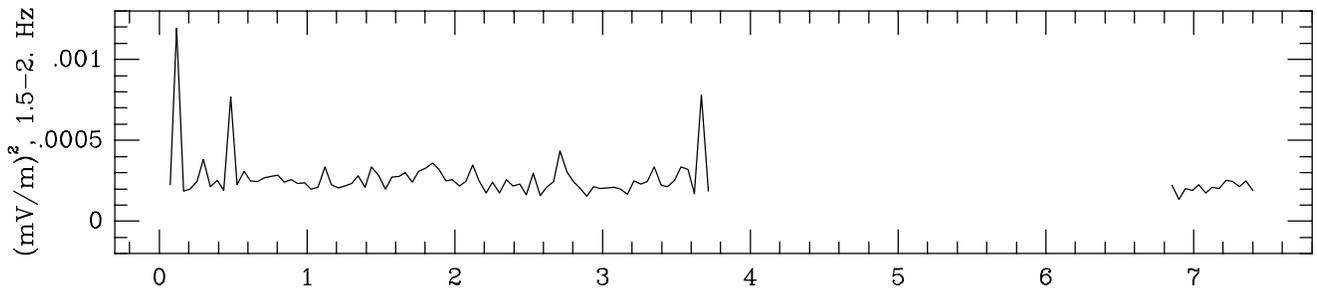
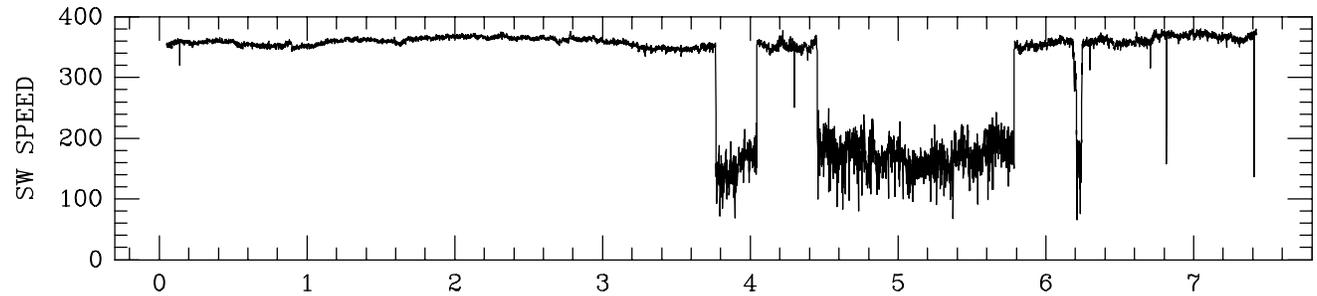
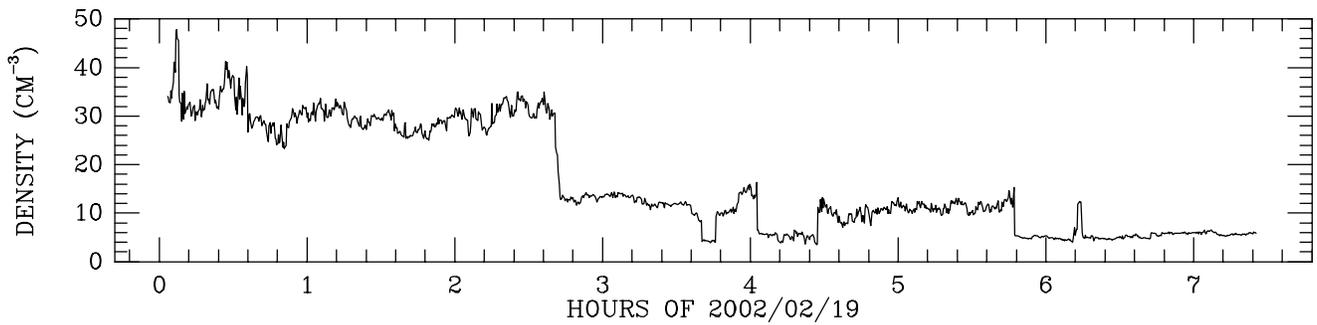

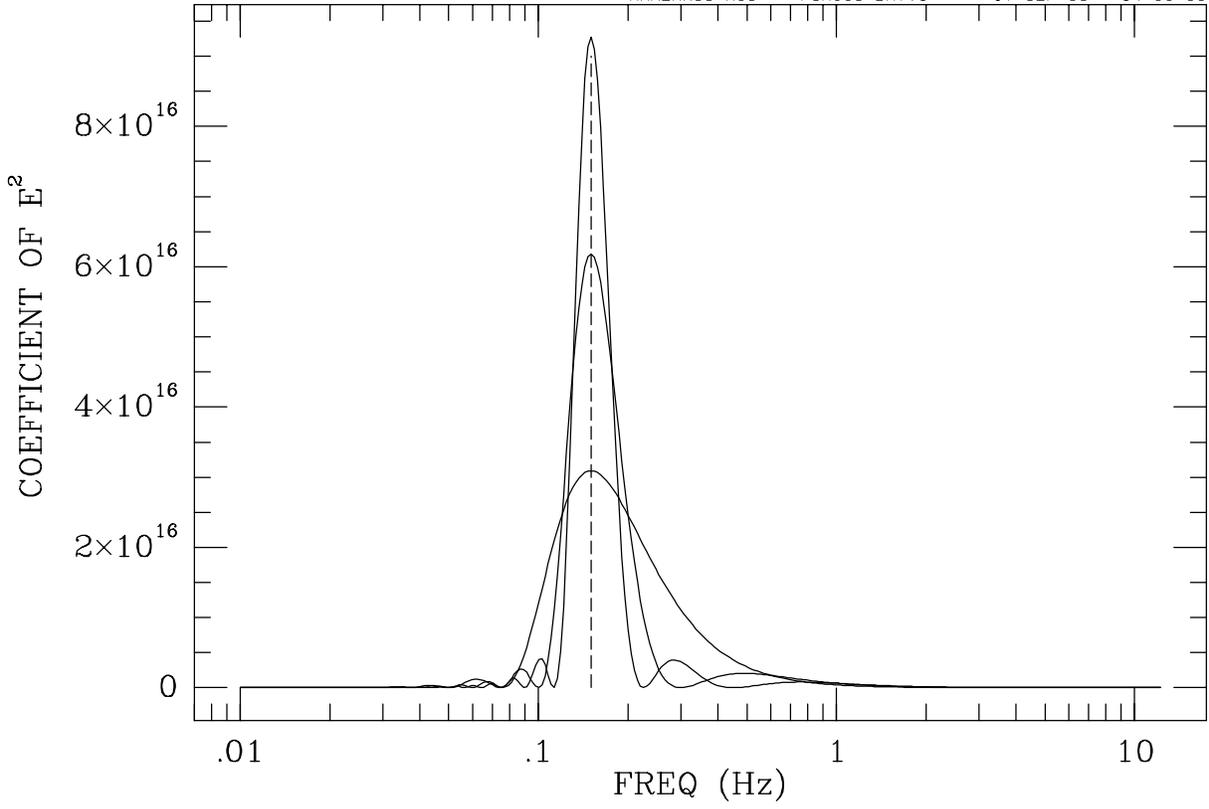
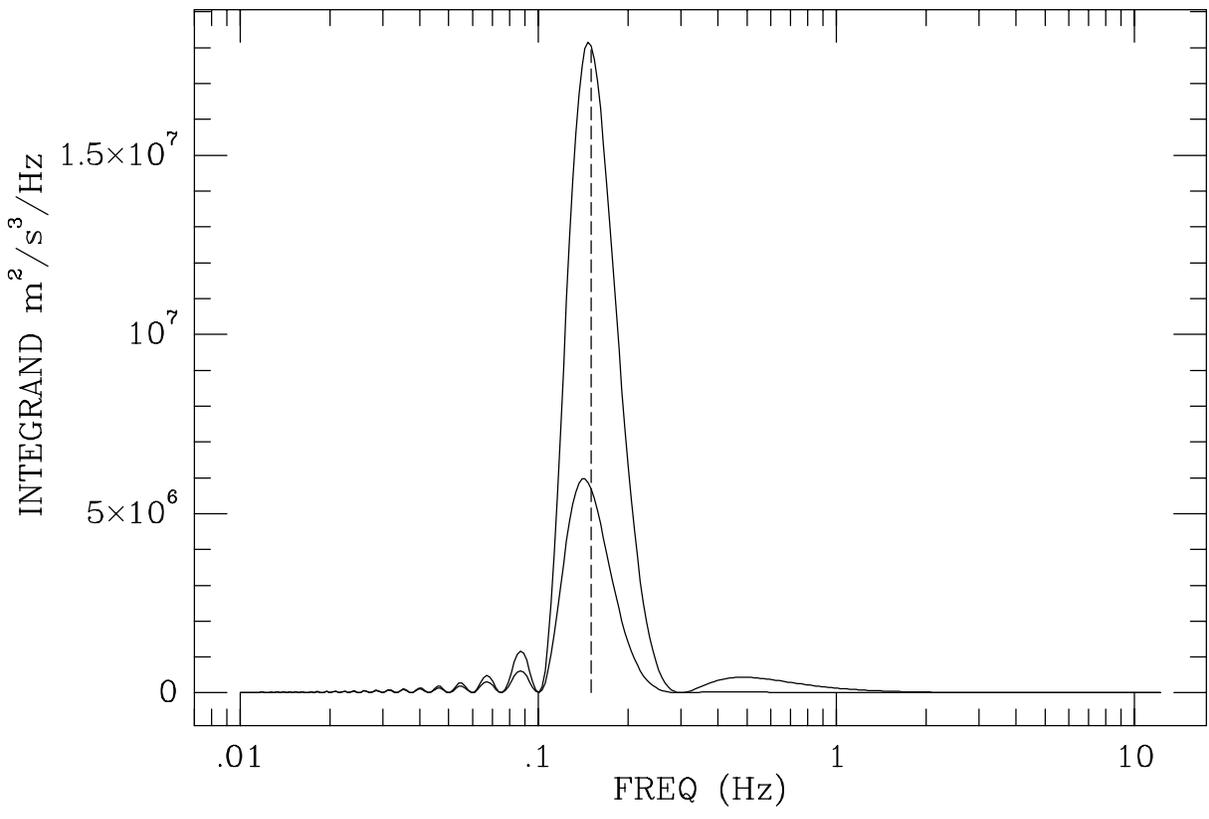